\documentclass[preprint,aps,a4paper]{revtex4-1}

\usepackage{mathrsfs}
\usepackage{graphicx}
\usepackage{multirow}
\usepackage{makecell}
\usepackage{booktabs}
\usepackage{lineno}

\begin{document}
 \title{Intrinsic Electronic Structure and Nodeless Superconducting Gap of $\mathrm{YBa_{2} Cu_{3} O_{7-\delta} }$ Observed by Spatially-Resolved Laser-Based Angle Resolved Photoemission Spectroscopy}
 
\author{Shuaishuai Li$^{1,2}$, Taimin Miao$^{1,2}$, Chaohui Yin$^{1,2}$, Yinghao Li$^{1,2}$,  Hongtao Yan$^{1,2}$,  Yiwen Chen$^{1,2}$, Bo Liang$^{1,2}$, Hao Chen$^{1,2}$, Wenpei Zhu$^{1,2}$,  Shenjin Zhang$^{3}$, Zhimin Wang$^{3}$, Fengfeng Zhang$^{3}$, Feng Yang$^{3}$, Qinjun Peng$^{3}$ , Chengtian Lin$^{4}$,  Hanqing Mao$^{1,2,5}$, Guodong Liu$^{1,2,5}$, Zuyan Xu$^{3}$,  Lin Zhao$^{1,2,5}$ and X. J. Zhou$^{1,2,5,*}$}
		
\affiliation{
		\\$^{1}$National Lab for Superconductivity, Beijing National Laboratory for Condensed Matter Physics, Institute of Physics, Chinese Academy of Sciences, Beijing 100190, China
		\\$^{2}$University of Chinese Academy of Sciences, Beijing 100049, China
		\\$^{3}$Technical Institute of Physics and Chemistry, Chinese Academy of Sciences, Beijing 100190, China
		\\$^{4}$Max Planck Institute for Solid State Research, Heisenbergstrasse 1, D-70569 Stuttgart, Germany
		\\$^{5}$Songshan Lake Materials Laboratory, Dongguan, Guangdong 523808, China
		\\$^{*}$Corresponding author: XJZhou@iphy.ac.cn
}

	\date{\today}
	
	\maketitle
	
	\newpage

{\bf The spatially-resolved laser-based high resolution ARPES measurements have been performed on the optimally-doped $\mathrm{\mathbf{YBa_{2}Cu_{3}O_{7-\delta}}}$ (Y123) superconductor. For the first time, we found the region from the cleaved surface that reveals clear bulk electronic properties. The intrinsic Fermi surface and band structures of Y123 are observed. The Fermi surface-dependent and momentum-dependent superconducting gap is determined which is nodeless and consistent with the d+$\mathit{i}$s gap form.}

      \vspace{3mm}


  The superconductivity mechanism of the high temperature cuprate superconductors remains to be one of the most prominent issues in condensed matter physics\cite{BKeimer2015}. The detection of intrinsic electronic structure and the determination of the superconducting gap symmetry are essential in understanding the unusual normal state properties and the superconductivity mechanism of the cuprate superconductors. Angle resolved photoemission spectroscopy (ARPES) has played a key role in studying the cuprate superconductors\cite{ADamascelli2003,JCCampuzano2004,XJZhou2007,JASobota2021}. However, most of the ARPES measurements are carried out on the Bi-based superconductors, particularly $\mathrm{Bi_{2} Sr_{2} CaCu_{2}O_{8} }$ (Bi2212), because it is easy to get smooth surface of the cleaved sample and measure the intrinsic electronic structure of the bulk material. Much less ARPES measurements have been performed on $\mathrm{YBa_{2}Cu_{3}O_{7-\delta}}$ (Y123), another prototypical cuprate superconductor with a $\mathrm{T_{c}}$ above the liquid nitrogen temperature\cite{JCCampuzano1990,JCCampuzano1991,RClaessen1991,GMante1991,RLiu1992,MLindroos1993,KGofron1993,MCSchabel1998A,MCSchabel1998B,DHLu2001,SVBorisenko2006,VBZabolotnyy2007,KNakayama2007,MAHossain2008,MOkawa2009,KNakayama2009,TDahm2009,DFournier2010,VBZabolotnyy2012,HIwasawa2018}. Because the cleaved surface of Y123 is polar which causes charge redistribution and surface self-doping, it is difficult to probe the intrinsic electronic structure of the bulk Y123\cite{JCCampuzano1990,RClaessen1991,RLiu1992,MCSchabel1998A,DHLu2001,VBZabolotnyy2007}. In particular, the Fermi surface-dependent and momentum-dependent superconducting gap has not been clearly determined in Y123\cite{KNakayama2009,MOkawa2009,DFournier2010}. The measurement of the intrinsic bulk electronic structure and superconducting gap in Y123 is important in establishing a general picture in understanding high temperature superconductivity in cuprate superconductors.

In the present paper, we report our observations of the intrinsic electronic structure and superconducting gap of Y123. By performing spatially-resolved laser-based ARPES, we successfully found the cleaved region where the CuO$_{2}$ planes are not self-doped and measured electronic structure and superconducting gap are intrinsic to the bulk Y123. The Fermi surface-dependent and momentum-dependent superconducting gap is clearly determined and a nodeless d-wave gap is observed in Y123.


 ARPES measurements were carried out using our lab-based laser ARPES systems equipped with the 6.994 eV vacuum-ultra-violet (VUV) laser and a DA30L hemispherical electron energy analyzer\cite{XJZhou2008GDLiu,WTZhang2018XJZhou}. The laser spot is focused to $\sim$15\,$\mu$m on the sample.  The energy resolution was set at 1\,meV and the angular resolution is $\sim$0.3$^{o}$, corresponding to a momentum resolution of $\sim$0.004 $\mathrm{\AA^{-1}}$.  High quality single crystals of $\mathrm{YBa_{2} Cu_{3} O_{7-\delta} }$ were grown by the self flux method\cite{CTLin1992}. The samples were post-annealed at $\mathrm{494\,^{o}C}$ under oxygen pressure of $\sim$1 atmospheres for 7 days. The obtained samples are  optimally doped with a $T_{c}$ of 92.7\,K and a transition width of $\sim$0.3\,K (Fig. 1c). The measured samples are twinned. All the samples were cleaved $in situ$ at a low temperature and measured in vacuum with a base pressure better than $5\times10^{-11}$ Torr at 17\,K. The Fermi level is referenced by measuring on a clean polycrystalline gold that is electrically connected to the sample.


  Through real-space point-by-point ARPES scanning measurements (Fig. S1 in Supplementary Materials), we found mainly four kinds of band structures on the Y123 cleavage surface, as shown in Fig. 1e. The Band Structure \uppercase\expandafter{\romannumeral1} (leftmost panel in Fig. 1e) has a broad band with a diffuse distribution of intensity in momentum space. It exhibits a clear energy gap at the Fermi level. The Band Structure \uppercase\expandafter{\romannumeral2} also has a broad band but without energy gap opening at the Fermi level. The Band Structure \uppercase\expandafter{\romannumeral3} shows two bands that are well separated in the momentum space. Both bands do not show energy gap opening at the Fermi level. The Band Structure \uppercase\expandafter{\romannumeral4} shows two sharp bands which are close in the momentum space. Both bands show energy gap opening at the Fermi level. Fig. 1b shows the spatial distribution of the Band Structure \uppercase\expandafter{\romannumeral3} and \uppercase\expandafter{\romannumeral4} on the cleaved sample surface. These two regions occupy only a small fraction ($<$5$\%$) of the surface area while the majority of the surface is occupied by the  Band Structure \uppercase\expandafter{\romannumeral1} and \uppercase\expandafter{\romannumeral2}.

  Figure 1a shows the crystal structure of Y123. It is usually considered that the cleaving occurs between the BaO layer and CuO chain layer, giving rise to two cleavage planes \uppercase\expandafter{\romannumeral1} and \uppercase\expandafter{\romannumeral2}\cite{HLEdwards1992,DJDerro2002,HIwasawa2018}. But only two kinds of cleavage planes can not account for the four kinds of band structures we have observed. Two more cleavage planes have to be considered that occur between the BaO layer and the CuO$_{2}$ layer, marked as cleavage plane \uppercase\expandafter{\romannumeral3} and \uppercase\expandafter{\romannumeral4} in Fig. 1a. In Fig. 1d, we show these four kinds of cleavage planes and analyse the charge distribution among different layers. It turns out that all these four cleavage planes are polar and,  according to the polar catastrophe model\cite{NNakagawa2006}, there are charge redistribution on the top layer. As a result, the CuO$_{2}$ planes are self-doped with the doping level increase from the original 0.16 to 0.33 in the cleavage plane \uppercase\expandafter{\romannumeral2} and \uppercase\expandafter{\romannumeral3}. This makes the top CuO$_{2}$ planes heavily overdoped and become nonsuperconducting. But the CuO$_{2}$ planes keep their doping level 0.16 in the cleavage plane \uppercase\expandafter{\romannumeral1} and \uppercase\expandafter{\romannumeral4}. This makes it possible to retain its intrinsic electronic structure and keep at the optimal doping level.

  By considering the charge distribution in different cleavage planes in Fig. 1d and the gap opening in the four kinds of band structures in Fig. 1e, and also considering the big (small) chance of cleavage for the cleavage plane \uppercase\expandafter{\romannumeral1} and \uppercase\expandafter{\romannumeral2} (\uppercase\expandafter{\romannumeral3} and \uppercase\expandafter{\romannumeral4}) in Fig. 1d and the large (small) occupation of the Band Structure \uppercase\expandafter{\romannumeral1} and \uppercase\expandafter{\romannumeral2} (Band Structure \uppercase\expandafter{\romannumeral3} and \uppercase\expandafter{\romannumeral4}), we can make a good correspondence between the four cleavage planes and the observed four kinds of band structures. In particular, the Band Structure \uppercase\expandafter{\romannumeral4} originates from the cleavage plane \uppercase\expandafter{\romannumeral4} where the top CuO$_{2}$ planes keep their original doping level 0.16. Both bands are sharp with clear gap opening. Therefore, the electronic structures are intrinsic to the bulk Y123. It is the first time that this kind of band structure has been observed in Y123 because its occupation area is rather small in the cleaved sample surface. We will focus on this band structure in the rest of the paper.

   Figure 2 shows the Fermi surface and band structures of Y123 measured at 17\,K from the red region in Fig. 1b. During the measurements, we overcome two technical issues. The first is the space charge effect caused by the small laser spot; we reduced the laser intensity to alleviate the space charge effect. The second is the light-induced modification effect; the measured area is modified by the light illumination over a period of time. To reduce the effect, we limited our acquisition time to finish the measurement on one region before the obvious sample modification occurs. These two issues put a strong limit on the data statistics. 
   
   Two bands are clearly observed that correspond to the bonding band (BB) and the antibonding band (AB) in Y123. The bonding band can be observed over a wide momentum space from the nodal to the antinodal regions. It becomes weaker when the momentum cut moves to the antinodal region. The gap opening at the Fermi level can be clearly observed on the bonding band which gets larger with the momentum cut moving from nodal to the antinodal regions. On the other hand, the antibonding band is clearly observed only around the nodal region; it becomes invisible when the momentum cut moves away from the nodal region. By considering the Fermi surface mapping in Fig. 2a and the corresponding band structures in Fig. 2c, and also referring to the Fermi surface measured from the previous ARPES measurements\cite{DFournier2010}, we arrived at the Fermi surface of the optimally-doped Y123 as plotted in Fig. 2a which consists of the bonding Fermi surface (BB), the antibonding Fermi surface (AB), and the chain Fermi surface (CH).  

   Figure 3 focuses on the band structure of Y123 along the nodal direction. Two sharp bands are clearly observed in Fig. 3a which corresponding to bilayer split bonding and antibonding bands. These two bands become more clearly observed in the second derivative image in Fig. 3b. The corresponding photoemission spectra (energy distribution curves, EDCs) show sharp superconducting coherence peaks as seen in Fig. 3c. The EDCs at the two Fermi momenta are plotted in Fig. 3d and their corresponding symmetrized EDCs are shown in Fig. 3e. Clear superconducting gap opening is observed on the two Fermi surface along the nodal direction.

   Figure 3f shows momentum distribution curves (MDCs) at different energies obtained from Fig. 3a. These MDCs can be well fitted by using two Lorentzians. The fitted dispersions and the MDC widths are plotted in Fig. 3g and Fig. 3h, respectively. The AB and BB bands exhibit clear kink in their dispersions (Fig. 3g), accompanied by the drop in their MDC width (Fig. 3h). To reveal the kink structure more clearly, in Fig. 3i, we plotted the energy difference between the measured dispersions and straight lines as shown in Fig. 3g. Well-defined peaks at $\sim$59\,meV are observed in Fig. 3i which corresponds to the kink position in the dispersions in Fig. 3g. The dispersion kink can be attributed to the electron-mode coupling. We note that, since the measured region keeps the original doping level, our measured kink structure is obviously stronger than those previously measured in self-doped Y123\cite{SVBorisenko2006,TDahm2009}.

   In Fig. 3j, we summarized the nodal bilayer splitting observed by the ARPES measurements on Y123 with different doping levels\cite{SVBorisenko2006, VBZabolotnyy2007,KNakayama2009,MOkawa2009,DFournier2010,VBZabolotnyy2012,HIwasawa2018}. Here the doping level has been corrected from their nominal values by considering the polar catastrophe model and the self-doping effect. The nodal bilayer splitting increases with the increasing doping level. This relation can be used to check on the real doping level of the measured region by measuring the nodal bilayer splitting. The nodal bilayer splitting in our measurement is the smallest. This further confirms that our measured region has the optimal doping level 0.16 and our measured electronic structures are intrinsic to the optimally-doped bulk Y123.  

   Now we come to the determination of the superconducting gap in Y123. Fig. 4a shows EDCs along the bonding Fermi surface. The corresponding symmetrized EDCs are plotted in Fig. 4b. These symmetrized EDCs can be fitted by the Norman formula\cite{MRNorman1998} to extract the superconducting gap. The obtained superconducting gap at different locations of the BB Fermi surface is plotted in Fig. 4f. Likewise, Fig. 4c shows EDCs along the antibonding Fermi surface and their corresponding symmetrized EDCs are presented in Fig. 4d. These symmetrized EDCs are also fitted by the Norman formula and the obtained superconducting gap for the AB Fermi surface is also plotted in Fig. 4f.

   As seen from Fig. 4f, the optimally-doped Y123 exhibits a superconducting gap ($\sim$7\,meV) even along the nodal direction. This is different from the other cuprate superconductors where the nodal superconducting gap is basically zero\cite{MHashimoto2014}. The presence of the s component in the superconducting gap of Y123 was also reported in the tunneling experiment\cite{JRKirtley2006}. The optimally-doped Y123 is different from other cuprate superconductors in that it has an orthorhombic crystal structure due to the presence of the CuO chain along the b direction. Whether the small anisotropy between a and b ($\sim$2\%) can cause such an obvious nodal gap needs further investigations.

   For a standard d-wave form, there is a zero gap along the nodal direction ($\theta$=45 degrees) (Fig. 4g). When there is an s component mixed in the d-wave form, there are two possibilities. One is the d+s form which still has gap node but the node position shifts away from $\theta$=45 degrees (Fig. 4h). When there are two domains present like in Y123, they will produce two kinds of gap form as shown in Fig. 4h. The other is the d+$\mathit{i}$s form where there is no longer gap node present (Fig. 4i). Our measured superconducting gap in Fig. 4f appears not consist with the d+s form because no gap node is observed. It is consistent with the d+$\mathit{i}$s form and the measured superconducting gap can be well fitted by the d+$\mathit{i}$s form, as shown in Fig. 4f.


   In summary, we have carried out spatially-resolved laser-based ARPES measurements on the optimally-doped Y123 superconductor. We found the region from the cleaved surface that represents the bulk electronic properties. The intrinsic Fermi surface and band structures of Y123 are obtained. Nodeless superconducting gap is observed which is consistent with the d+is gap form.

    \vspace{3mm}

    \noindent {\bf Acknowledgement} This work is supported by the National Natural Science Foundation of China (Grant Nos. 11888101 and 11974404), the National Key Research and Development Program of China (Grant Nos. 2021YFA1401800 and 2018YFA0704200), the Strategic Priority Research Program (B) of the Chinese Academy of Sciences (Grant No. XDB25000000 and XDB33000000), the Youth Innovation Promotion Association of CAS (Grant No. Y2021006), Innovation Program for Quantum Science and Technology (Grant No. 2021ZD0301800) and the Synergetic Extreme Condition User Facility (SECUF).

    \vspace{3mm}

    \noindent {\bf Author Contributions}\\
    X.J.Z. and S.S.L. proposed and designed the research. S.S.L., H.T.Y. and Y.W.C carried out the ARPES experiments. C.T.L grew the single crystals.  T.M.M., C.H.Y., Y.H.L., B.L., H.C., W.P.Z., S.J.Z., Z.M.W., F.F.Z., F.Y., Q.J.P., G.D.L., H.Q.M., L.Z., Z.Y.X. and X.J.Z. contributed to the development and maintenance of Laser-ARPES systems. S.S.L. and X.J.Z. analyzed the data and wrote the paper. All authors participated in discussions and comments on the paper.

    \newpage

    \begin{figure*}[tpb]
    \begin{center}
    	\includegraphics[width=1.0\columnwidth,angle=0]{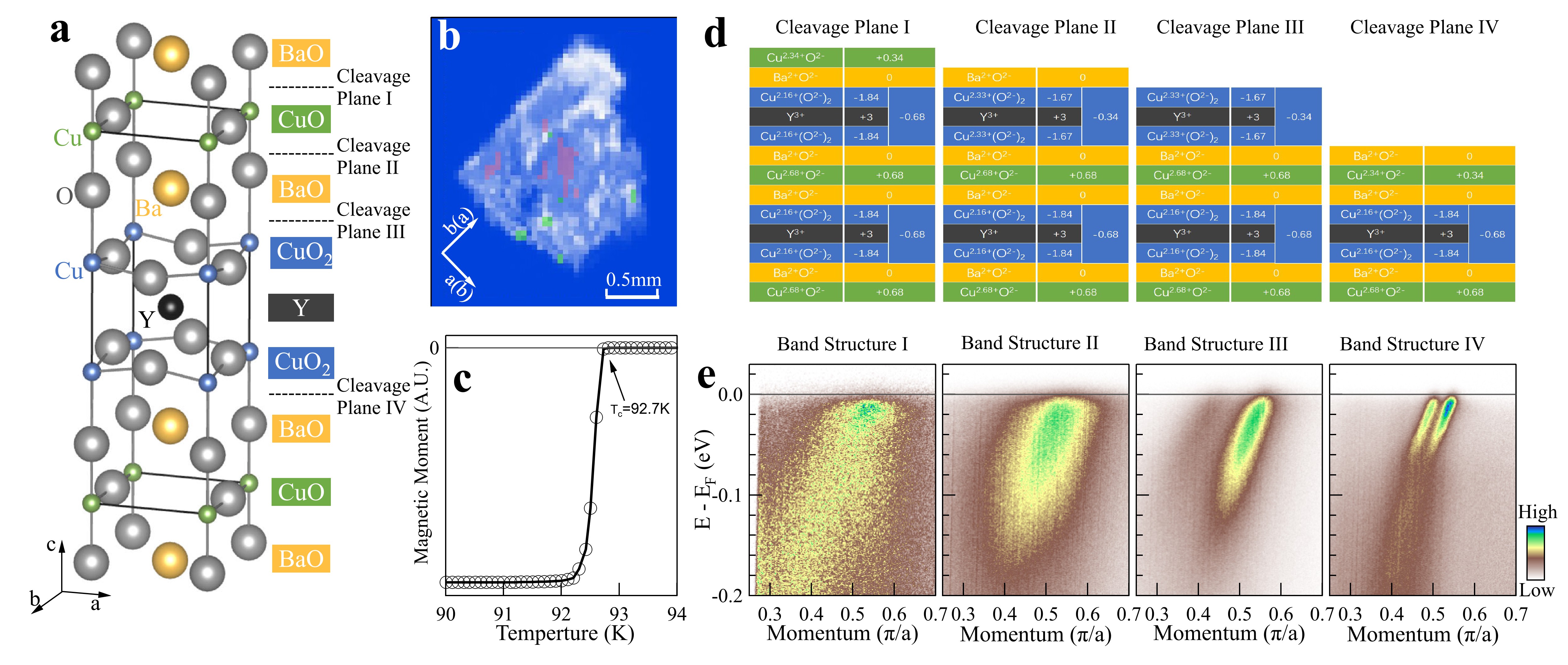}
    \end{center}

    \caption{{\bf Spatially resolved ARPES measurements on different cleavage surfaces of Y123. } (a) Crystal structure and four possible cleavage surfaces of Y123. (b) Spatial distribution of spectral intensity on the cleaved sample surface obtained by real-space point-by-point scanning of ARPES measurements. (c) Magnetic measurement of the superconducting transition temperature of the Y123 sample with a magnetic field of 2\,Oe. (d) Schematic four kinds of cleavage planes and their charge distribution analyses. In each panel, the left half shows the atom distribution and their valence states in each layer while the right half shows the overall charge in the layer(s). (e) Four main  band structures observed in Y123, which are attributed to the  four cleavage planes in (d). The region where the Band Structure \uppercase\expandafter{\romannumeral3} is observed is marked in green in (b) while the region where the Band Structure \uppercase\expandafter{\romannumeral4} is observed is marked in red in (b).}
    \end{figure*}

    \begin{figure*}[tbp]
	\begin{center}
    \includegraphics[width=1.0\columnwidth,angle=0]{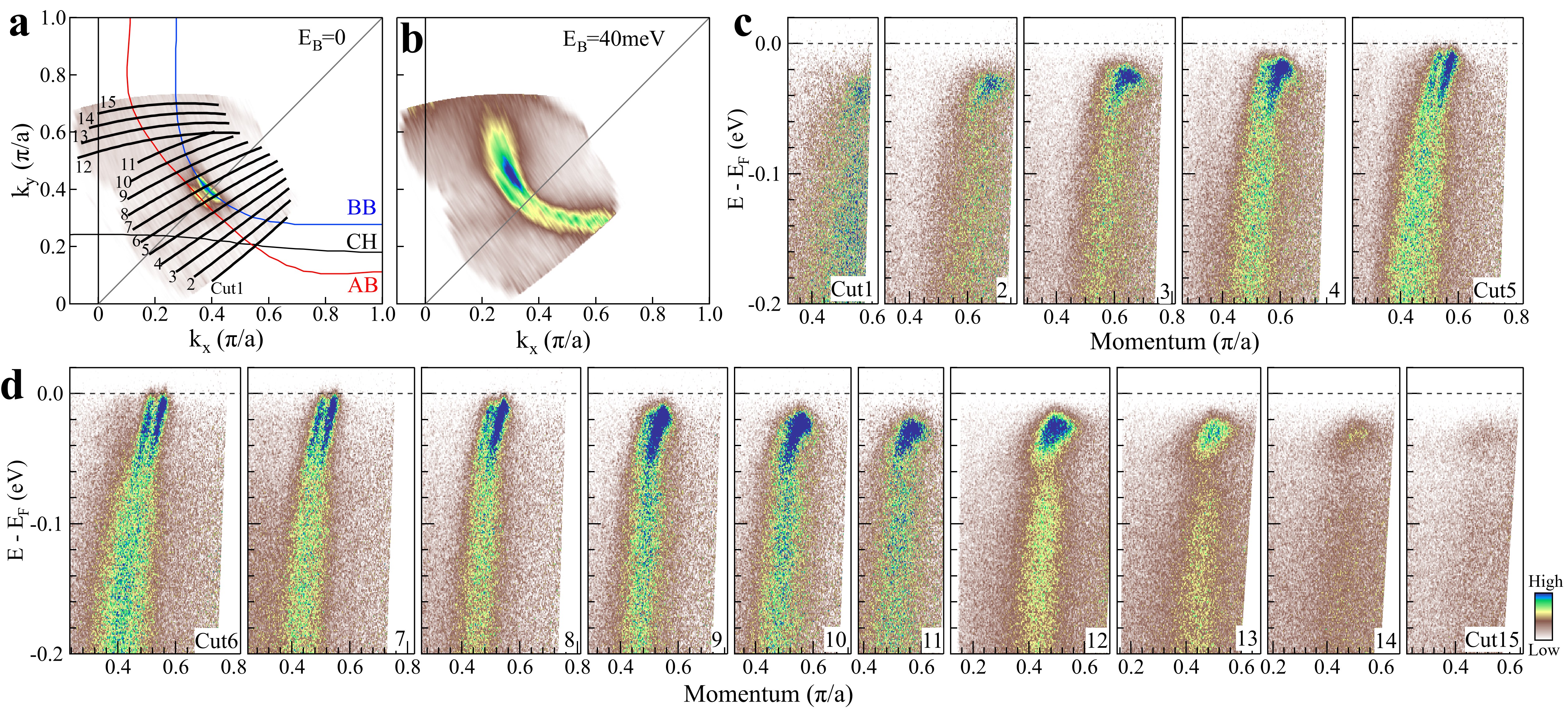}
	\end{center}
	\caption{{\bf Intrinsic electronic structure of Y123 measured at 17\,K in the superconducting state from the red region in Fig. 1b.} (a) Measured Fermi surface mapping. Three Fermi surface sheets from the bonding band (BB), the antibonding band (AB) and the CuO chain band (CH) are plotted by considering our measurements and the previous ARPES measurements\cite{DFournier2010}. (b) Constant energy contour at the binding energy of 40\,meV. (c,d) Band structures along different momentum cuts. The location of the momentum cuts is marked by the black lines in (a). }
	\end{figure*}

  \begin{figure*}[tbp]
	\begin{center}
\includegraphics[width=1.0\columnwidth,angle=0]{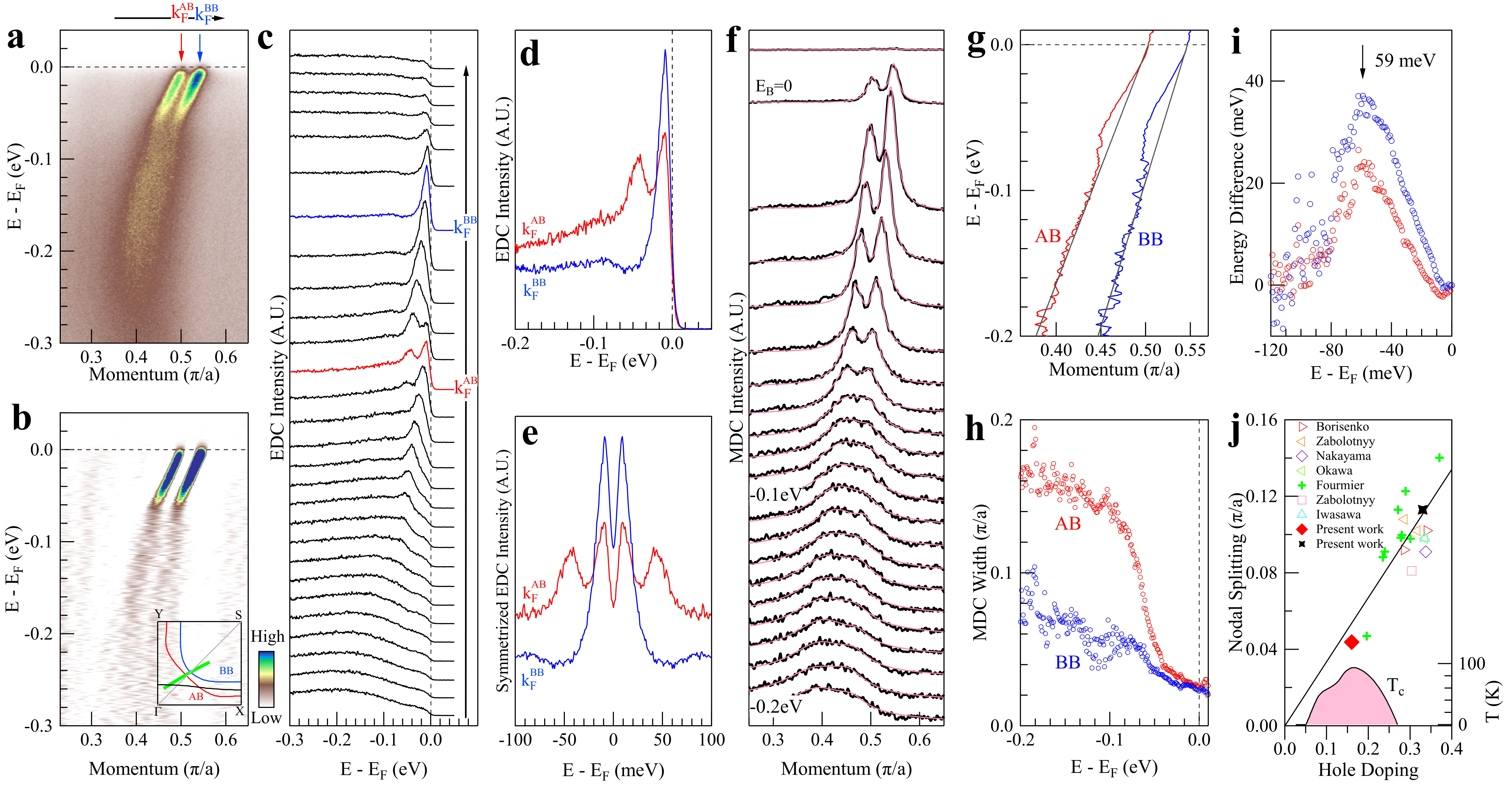}
	\end{center}
	
	 \caption{{\bf Intrinsic bilayer splitting along the nodal direction in Y123.} (a) Band structure measured at 17K along the nodal direction. The location of the momentum cut is marked by the green line in the inset of (b). (b) The corresponding MDC second derivative image of (a). (c) The corresponding EDCs of (a). (d) EDCs at the Fermi momenta of the bonding band (k$_{F}^{BB}$) and the antibonding band (k$_{F}^{AB}$). (e) Corresponding symmetrized EDCs  from (d). (f) MDCs at different energies obtained from (a). The MDCs are fitted by two Lorentzians as plotted by red lines. (g) Dispersions of the BB and AB bands obtained from fitting the MDCs at different energies as shown in (f). (h) The MDC width (full width at half maximum, FWHM) as a function of energy for the BB and AB bands obtained from MDC fitting in (f). (i) The energy difference between the measured BB and AB dispersions and the straight lines plotted in (g). (j) The momentum difference of the nodal bilayer splitting as a function of doping level in Y123 summarized from the previous ARPES measurements\cite{SVBorisenko2006, VBZabolotnyy2007,KNakayama2009,MOkawa2009,DFournier2010,VBZabolotnyy2012,HIwasawa2018} and the present study. For the previous measurements, the doping level is derived from the nominal doping of the bulk sample and corrected by considering the polar catastrophe model.}

	\end{figure*}

    \begin{figure*}[tbp]
	\begin{center}
		\includegraphics[width=1.0\columnwidth,angle=0]{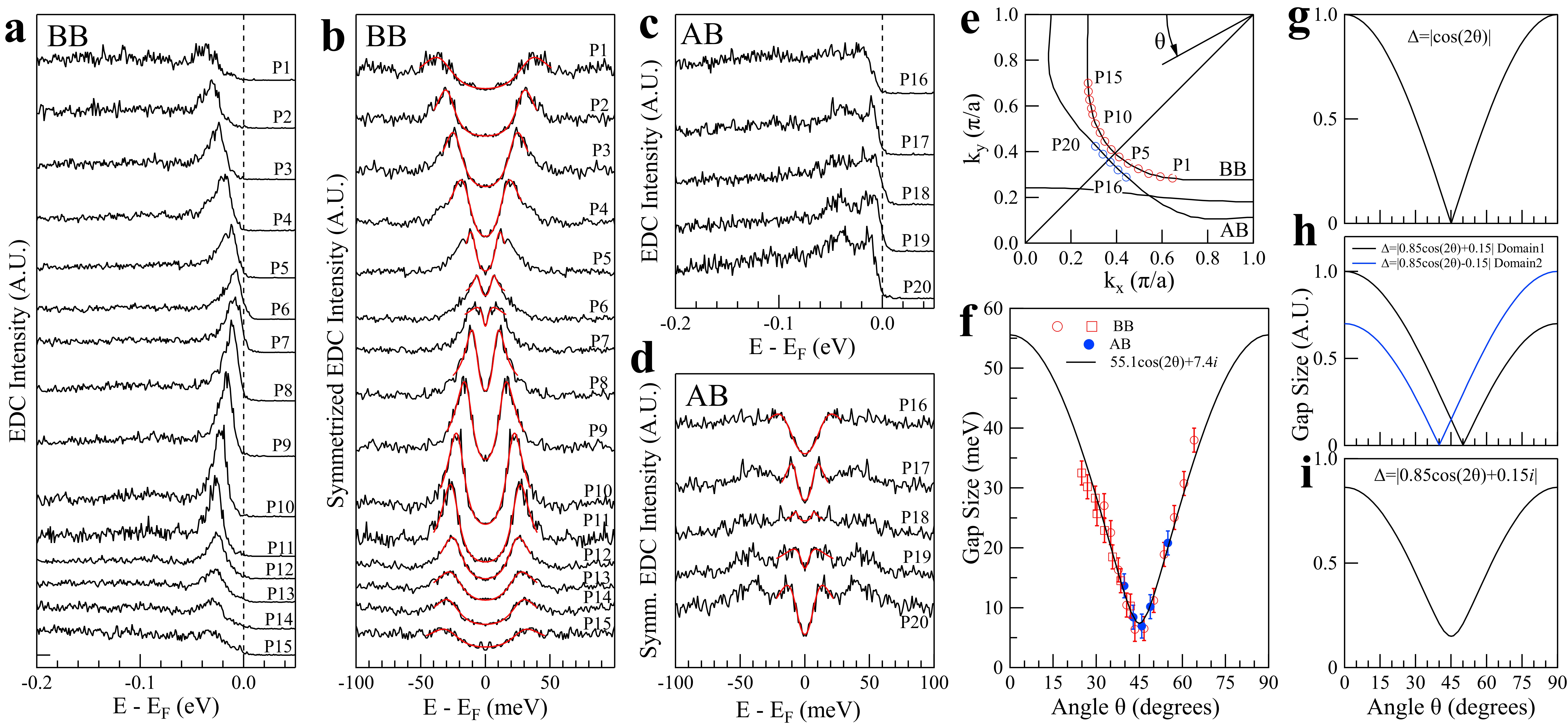}
	\end{center}
	
	\caption{{\bf Superconducting gap of Y123 measured at 17\,K.} (a) EDCs along the BB Fermi surface. The location of the momentum points is marked by red circles in (e). (b) The corresponding symmetrized EDCs from (a). These curves are fitted by the Norman formula as plotted by the red lines. (c) EDCs along the AB Fermi surface. The location of the momentum points is marked by blue circles in (e). (d) The corresponding symmetrized EDCs from (c). These curves are fitted by the Norman formula as plotted by the red lines. (e)  Schematic Fermi surface of Y123 with the Fermi momentum points marked. (f) Momentum-dependent superconducting gap of the BB Fermi surface (red symbols) and the AB Fermi surface (blue circles). The black line represents a fitting with the d+$\mathit{i}$s wave. (g) Momentum-dependent superconducting gap in a standard d-wave form. (h) Momentum-dependent superconducting gaps in a d+s form for the two perpendicular domains in Y123. (i) Momentum-dependent superconducting gap in the d+$\mathit{i}$s form.}

	\end{figure*}

\end{document}